\documentclass[twoside]{article}

\usepackage[sc]{mathpazo} 
\usepackage[T1]{fontenc} 
\usepackage{microtype} 

\usepackage[hmarginratio=1:1,top=32mm,columnsep=20pt]{geometry} 
\usepackage{multicol} 
\usepackage{hyperref} 

\usepackage[hang, small,labelfont=bf,up,textfont=it,up]{caption} 
\usepackage{booktabs} 
\usepackage{float} 

\usepackage{lettrine} 
\usepackage{paralist} 

\usepackage{enumitem}
\setlist[enumerate]{leftmargin=*}

\usepackage{fancyhdr} 
\usepackage{url}
\usepackage{xspace}
\newcommand{\eg}{e.g.,\xspace}

\title{\vspace{-15mm}\fontsize{24pt}{10pt}\selectfont\textbf{
A Collaborative Approach to Computational Reproducibility
}} 

\author{Fernando Chirigati\\
Reproducibility Editor, Information systems\\
New York University\\
\and
Rebecca Capone\\
Publisher\\
Elsevier Inc\\
\and
Dennis Shasha\\
Editor-in-Chief, Information systems\\
New York University\\
\and
R\'{e}mi Rampin\\
New York University\\
\and
Juliana Freire\\
New York University\\
}
\date{}

\begin{document}

\topskip0pt
\vspace*{\fill}
\textbf{This is an electronic post-print version of an article published in Information
Systems, Volume 59, July 2016, Pages 95-97, ISSN 0306-4379.}\\
The Journal of Information Systems is available online at:\\
\url{http://www.sciencedirect.com/science/journal/03064379}.\\
URL to published article: \url{https://doi.org/10.1016/j.is.2016.03.002}.
\vspace*{\fill}

\maketitle 

\thispagestyle{fancy} 


\begin{multicols}{2} 

\lettrine[lraise=0.1, nindent=0.2em, slope=-.5em]{R}{eproducing} and verifying experiments contributes to science in several ways. First, reproducibility enables reviewers (and readers in general) to verify the outcomes presented in papers, which is crucial for science to be self-correcting. Second, it allows new methods to be objectively compared against methods presented in previous publications. Third, studies indicate that reproducibility increases impact, visibility, and research quality~\cite{begley@nature2012,brody@thesis2006,hitchcock@2009,kovacevic@2007,lawrence@nature2001,piwowar@plos2007,vandewalle@ieee2009}, and helps defeat self-deception~\cite{nuzzo@nature2015}. Finally, computational reproducibility, by creating an executable artifact, enables researchers to build on top of previous work directly by simply extending the software.

Although a standard in natural science, reproducibility has been only episodically applied in experimental computer science. Scientific papers often present a large number of tables, plots and pictures that summarize the obtained results, but then loosely describe the steps taken to derive them. Not only can the methods and the implementation be complex, but also their configuration may require setting many parameters and/or depend on particular system configurations. As a consequence, reproducing the results from scratch is time-consuming, error-prone, and sometimes just infeasible. This has led to a credibility crisis in computational science~\cite{donoho@cise2009}. Previous studies showed that the fraction of papers in various computer science conferences that can be effectively reproduced is discouragingly low~\cite{bonnet@2011,collberg@tr2015,kovacevic@2007,vandewalle@ieee2009}.

While many researchers recognize the importance of reproducibility, they are often held back by the challenge of making it happen. Authors must describe and encapsulate the entire experiment, which includes data, parameters, source code, dependencies and environment, so that the results can be properly verified and explored. If the experiment has not been systematically documented and made reproducible from the start, it may be hard to track all the necessary components to include in such compendium, and important aspects may be mistakenly left out. As an example, some numerical models, if not fully described, may lead to different implementations that are mathematically equivalent, but numerically different, rendering them irreproducible~\cite{crook@springer2013}. Even when the original researchers have tried to make their results reproducible, follow-on researchers may not be able to reproduce the results for several reasons: insufficient documentation; the experiment may not run in their operating system; there may be missing libraries; library versions may be different; and the inability to install all the required dependencies.

The practical difficulties that reproducibility involves end up overshadowing its clear benefits to science. The process is often seen as a burden to both authors and reviewers. In fact, different surveys from various domains indicate that the effort and time required to make an experiment reproducible is one of the main reasons why authors do not do so~\cite{bonnet@2011,collberg@tr2015,tenopir@plos2011}.

Overall, to increase the practice of reproducibility in computational science, we have identified two main goals:
\begin{enumerate}
\item \emph{Usability:} development of tools that make it easier and significantly less time-consuming for authors to do reproducible research, and for reviewers to execute computational artifacts (and modify them) corresponding to published results.
\item \emph{Incentives:} a new publication model that recognizes the efforts of making experiments reproducible (for authors) and verifying published scientific results (for reviewers).
\end{enumerate}

Fortunately, a plethora of reproducibility solutions that address the first goal have been recently designed and implemented by the community. In particular, packaging tools (\eg ReproZip\footnote{\url{https://vida-nyu.github.io/reprozip/}}) and virtualization tools (\eg Docker\footnote{\url{https://www.docker.com/}}) are promising solutions towards facilitating reproducibility for both authors and reviewers. For example, with ReproZip~\cite{chirigati@sigmod2016}, authors can automatically track the dependencies of their existing experiments and create a self-contained package using only two commands. Reviewers can then use ReproZip to unpack, reproduce, and vary experiments using as few as three commands, even in a system that differs from the operating system of the authors' original environment. Docker~\cite{boettiger@sigops2015} complements ReproZip in the sense that it is able to create a virtualization environment for the experiment that is lightweight and easy to use and deploy.

To address the second goal, we have implemented a new publication model for the Reproducibility Section of Information Systems Journal. In this section, \emph{authors submit a reproducibility paper} that explains in detail the computational assets from a previous published manuscript in Information Systems. Submission is by invitation only.

A reproducibility paper describes all the software and data used to derive the published results, as well as provides instructions on how to reproduce and validate such results. In addition, authors can use this paper to discuss the benefits and challenges they encountered in making their experiment reproducible. Using Mendeley Data\footnote{\url{https://data.mendeley.com/}}, authors also submit their code, data, and optionally a ReproZip package or a Docker container to make the review process easier. Reviewers not only review the reproducibility paper, but also validate the results and claims published in the original manuscript. Once the paper is accepted, \emph{reviewers also become co-authors} and are encouraged to add a section in the paper that states the extent to which the software is portable, is robust to changes, and is likely to be usable as a subcomponent or as a basis for comparison by future researchers. The review is not blinded, so authors and reviewers are encouraged to engage in a discussion about the validity of the experimental results as many times as necessary.

This model addresses our second goal in two ways. First, it incentivizes authors to make their experiment reproducible by creating a second publication from their effort. Second, it recognizes the difficulties of the review's job---which, even when using ReproZip or Docker, may be time-consuming---thus allowing reviewers to become co-authors of the same publication.

Our first reproducibility paper was published online in January 2016~\cite{Wolke2016}. It validates the results and claims presented in the authors' original manuscript~\cite{Wolke2015}. The experimental setup in this paper is complex, involving many details and configuration parameters. Using the manuscript alone, it would not be possible to reproduce its results. The corresponding reproducibility report, together with the published computational assets (which were derived using ReproZip and Docker), greatly increases the likelihood that readers will reproduce the results and reuse the approach in future research.

This model is also the latest effort in Elsevier's history of exploring the potential for reproducibility in scientific publications in computer science and other disciplines. Beginning with the Executable Paper Grand Challenge in 2010, Elsevier has piloted several models of reproducibility in a variety of disciplines with data-rich science. The model here, \emph{Invited Reproducibility Reports}, is an article type that joins a new class of scientific publication, creating publishable and citable artifacts to accompany reported research experiments.

The benefits of the Reproducibility Report are threefold: proven experimental reproducibility for researchers, a collaborative full-fledged academic publication for the original authors and reproducibility reviewers, and a canonical reference point for the vetted experimental components. Further, by using the existing infrastructure of Mendeley Data and GitHub, and user-friendly tools such as ReproZip and Docker, reproducibility reports constitute a model that is easily adoptable by other scientific journals, making reproducibility a reality in computational experiments.

We believe this new publication model will improve the degree of reproducibility in all computational sciences, thus increasing the reliability and usefulness of scientific research.

\bibliographystyle{splncs03}
\bibliography{editorial}

\end{multicols}

\end{document}